# Association-sensory spatiotemporal hierarchy and functional gradient-regularised recurrent neural network with implications for schizophrenia

Subati Abulikemu[1,2*], Puria Radmard[3], Michail Mamalakis[1,2,4], John Suckling[1,2]

1. Department of Psychiatry, University of Cambridge, Cambridge, UK
2. Centre for Human-Inspired Artificial Intelligence, University of Cambridge, Cambridge, UK
3. Department of Engineering, University of Cambridge, Cambridge, UK
4. Department of Computer Science and Technology, University of Cambridge, Cambridge, UK

\* Corresponding author

Email: ss2905@cam.ac.uk

**Abstract**

The human neocortex is functionally organised at its highest level along a continuous sensory-to-association (AS) hierarchy. This study characterises the AS hierarchy of patients with schizophrenia in a comparison with controls. Using a large fMRI dataset (N=355), we extracted individual AS gradients via spectral analysis of brain connectivity, quantified hierarchical specialisation by gradient spread, and related this spread with connectivity geometry. We found that schizophrenia compresses the AS hierarchy indicating reduced functional differentiation. By modelling neural timescale with the Ornstein-Uhlenbeck process, we observed that the most specialised, locally cohesive regions at the gradient extremes exhibit dynamics with a longer time constant, an effect that is attenuated in schizophrenia. To study computation, we used the gradients to regularise subject-specific recurrent neural networks (RNNs) trained on working memory tasks. Networks endowed with greater gradient spread learned more efficiently, plateaued at lower task loss, and maintained stronger alignment to the prescribed AS hierarchical geometry. Fixed point linearisation showed that high-range networks settled into more stable neural states during memory delay, evidenced by lower energy and smaller maximal Jacobian eigenvalues. This gradient-regularised RNN framework therefore links large-scale cortical architecture with fixed point stability, providing a mechanistic account of how gradient de-differentiation could destabilise neural computations in schizophrenia, convergently supported by empirical timescale flattening and model-based evidence of less stable fixed points.

**Introduction**

The human neocortex operates through coordinated, hierarchically organised modules that support both integrative and specialised functions. In schizophrenia, this architecture is posited to be disrupted by dysconnectivity—aberrant connections among neural ensembles that impair information processing[1–3]. Recent cortical mapping revealed a continuous hierarchy extending from primary sensory systems to transmodal association networks, evident across scales from gene expression and cytoarchitecture to morphology and macroscale functional connectivity (FC)[4–8]. Rather than describing FC as compartmentalised pairwise links, the principal association-sensory (AS) functional gradient delineates a smooth, computationally meaningful transition from perception to abstract cognition[9]. Here, we leverage this gradient framework to explore dysconnectivity in schizophrenia through two complementary approaches. First, we experimentally examine how the AS gradient reorganises in schizophrenia and maps onto intrinsic neural timescales. Second, we embed empirically derived gradients as architectural constraints in recurrent neural networks (RNNs), allowing us to mechanistically probe how hierarchical disruptions degrade cognitive computations.

Functional dysconnectivity holds promise as an explanatory framework for schizophrenia, supported by extensive fMRI evidence of abnormal FC, reduced small-worldness, and diminished functional segregation[10–12]. However, regional connectomic findings remain inconsistent, reflecting the idiosyncratic nature of brain organisation compounded by heterogeneity in cortical mapping strategies. Network-level approaches often enforce spatial independence and arbitrary thresholds, overlooking the brain's continuous functional landscape[13]. From a computational standpoint, capturing global hierarchical organisation is critical for mechanistically modelling how network architecture shapes information diffusion, neural dynamics stability, and cognition. Conceptualising dysconnectivity through continuous, low-dimensional FC gradients, whose spatial spreads encode hierarchical specialisation, thus offers a robust alternative. Indeed, recent studies have consistently reported compression and functional de-differentiation of the AS gradient in schizophrenia[14–16]. Seeking an explanation of how gradient alterations link to dynamics and cognitive computations, we begin by examining the relationship between spatial AS gradient maps and the brain's temporal organisation.

The hierarchical structure inherent in the neocortex's functional architecture is likely mirrored in its dynamics, specifically its neural timescales[17,18]. Multimodal evidence suggests a global hierarchy of temporal integration windows, indexed by signal autocorrelation decay, which lengthens from early sensory to higher-order areas[19–23]. Within this global trend, more granular system-specific temporal gradients have been indicated[24,25]. However, the relationship between the AS gradient and temporal integration may not be strictly monotonic, considering the modular organisation of neural systems and evidence connecting longer timescales to greater within-community

FC[26]. Under the gradient framework, AS gradient extremes encode highly specialised nodes with functionally similar neighbours. Consequently, protraction toward either extreme of the AS gradient likely reflects increasing within-subsystem integration. We therefore hypothesise that the spatiotemporal mapping reflects nested timescale hierarchies within each subsystem.

The neurodynamical hypothesis of schizophrenia proposes that brain network activity is destabilised by shallow attractor states[27,28]. These attractors represent stable patterns of neural activity that underpin cognitive processes such as working memory. Computationally, instability implies that neural networks fail to robustly maintain these patterns and are easily perturbed, causing sudden and more frequent transitions across states. This fragility aligns with a dampening of neural integration windows in schizophrenia, leading to temporal fragmentation in neural coding. Indeed, resting-state fMRI studies have evinced brain-level timescale reductions in schizophrenia relative to controls[24,29]. Alterations in the intrinsic relationship between timescales and the AS gradient offer an explanatory framework for differing functional connectivity in schizophrenia, and how it translates to characteristic effects on cognition.

Moving beyond empirical associations to understand how gradient disruptions influence cognitive computations necessitates computationally explicit modelling. To this end, RNNs serve as powerful means for generating and testing mechanistic hypotheses of neural computations[30–35]. When task-optimised RNNs are trained on the same behavioural paradigms used during neural recordings, they can reproduce observed population dynamics and reveal previously unknown computational mechanisms[32]. The mapping from network architecture to behaviour is typically many-to-one, such that substantially different connectivity patterns could achieve comparable performance, thereby highlighting the role of regularisations in the optimisation process[30,36]. In a complementary line of research, empirical connectivity can be embedded as biologically informed organisational constraints on the network during training[37–40]. From the resulting nonlinear dynamical systems, one can examine how attractor-like, slow regions in the state space implement cognitive computations; for instance, through linearisation analyses of emergent dynamical motifs[34,41]. Under this framework, we explicitly probe the computational and dynamical properties associated with the AS gradient, hypothesised to be compressed in schizophrenia, by training empirical gradient-regularised RNNs on cognitive tasks.

In this study, we aimed to elucidate the implications of the degree of differentiation along the AS gradient—quantified as the gradient range—for neural dynamics and computation. We approached this by integrating experimental findings from resting-state fMRI with theoretical models. First, we extracted the AS gradient and assessed its intrinsic mapping to neural timescales, with a particular focus on potential alterations in schizophrenia. We sought to replicate previously reported AS gradient de-differentiation and test whether schizophrenia exhibits a dampened, more homogeneous distribution of neural integration windows, with diminished relative slowness in specialised, locally cohesive communities. Beyond these empirical studies, we developed a generative framework that embeds the empirically derived AS gradient as connectivity constraints in RNNs, leveraging the link between gradient spread and connectivity weight geometry. We specifically focused on the AS gradient as it encapsulates the most global hierarchical specialisation, covering both lower- and higher-order systems. By regularising the working memory-performing RNNs with AS connectivity matrices generated from empirical AS gradients, we directly probe how variations in AS gradient spread modulate network learning and dynamical stability.

## Results

### *Experimental*

<u>Principal cortical gradient and its reorganisation in schizophrenia</u>

We analysed resting-state functional MRI (fMRI) data from the Bipolar and Schizophrenia Network for Intermediate Phenotypes (BSNIP) consortium, comprising 186 healthy controls and 169 schizophrenia proband patients across four study sites (Baltimore N=132, Hartford N=92, Dallas N=81, and Boston N=43) [42,43]. Our goal was to investigate whether schizophrenia involves disruptions in hierarchical functional specialisation. To this end, we decomposed the sparsified and similarity transformed functional connectivity (FC) matrices through diffusion map embedding (Fig 1A & B; see Methods: Spatial gradients via diffusion map). We specifically focused on the principal association-sensory (AS) gradient, $\psi_{AS}$—the lowest-frequency (most global) eigenmode capturing the unimodal sensory to transmodal association continuum. The following results were consistent across different network density thresholds applied at sparsification (Supplementary Fig 1); the reported results are based on retaining 50 edges per node (14% of connections per region given 360 cortical regions; see Methods: Spatial gradients via diffusion map). All subsequent regression analyses use z-scored continuous variables.

The information explanatory power of $\psi_{AS}$ was significantly higher in healthy controls (mean eigenvalue ratio = .21) than in schizophrenia patients (mean eigenvalue ratio = .19), after adjusting for age, gender, site, and TR using multiple regression analysis ($\beta = .45, p < .001, 95\% CI = [.24, .66]$). A critical focus of this study was the gradient range, indexing the extent of FC differentiation and hierarchical organisation along the eigenmode. Using the same regression approach, we observed a significantly higher range of $\psi_{AS}$ in healthy controls (mean $range_{AS}$ = .22) compared to schizophrenia patients (mean $range_{AS}$ = .20; $\beta = .35, p < .001, CI = [.14, .55]$; Fig 1C). In short, functional specialisation along the global AS hierarchy is attenuated in schizophrenia.

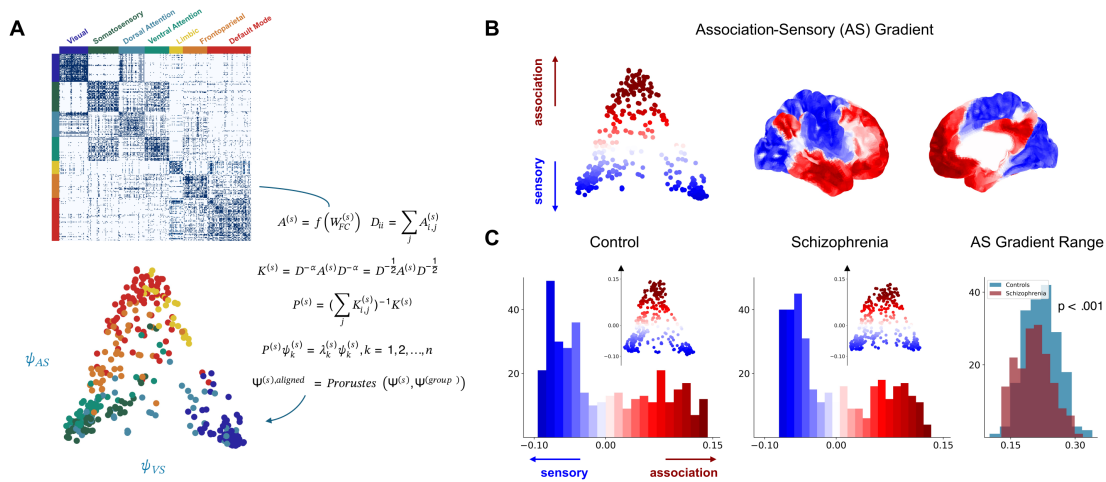

**Fig 1 Compressed association-sensory gradient in schizophrenia. (A)** Diffusion map embedding was applied to the sparsified and similarity-transformed FC matrix, producing a spectral decompositions of the similarity matrix. The two primary gradients are the AS gradient ($\psi_{AS}$) and the unimodal gradient differentiating sensory modalities ($\psi_{VS}$). **(B)** Cortical projection of the principal AS gradient illustrates the continuous transition from unimodal (blue) to transmodal regions (red). **(C)** The approximate bimodal distribution of the AS gradient values shows a significant compression in schizophrenia.

To evaluate the behavioural relevance of the altered $\psi_{AS}$ eigenmode spread, we examined its association with cognitive function measured by the composite Brief Assessment of Cognition in Schizophrenia (BACS), while accounting for diagnosis and the confounding variables above. Regression analyses revealed that $range_{AS}$ was significantly and positively associated with cognition ($\beta = .11$, $p = .03$, $CI = [.01, .21]$; $N_{HC} = 151$, $N_{SZ} = 151$) across all samples. However, within the schizophrenia cohort alone, this relationship did not remain significant ($\beta = .14$, $p = .09$, $CI = [-.02, -.29]$). Within the schizophrenia group, $range_{AS}$ showed a significant negative association with positive symptom severity (PANSS Positive Scale; $\beta = -.19$, $p = .02$, $CI = [-.35, -.03]$; $N_{SZ} = 159$), but no significant relationship with negative symptom severity ($\beta = -.06$, $p = .52$, $CI = [-.23, .12]$).

Eigenmode spread and network specialisation

Here we quantify, at the subject level, how the spread of AS gradient, $\psi_{AS}$, is reflected in the weight organisation of the connectivity matrix, $W_{FC}$; this empirical relationship forms the basis for generating the AS weight matrix used later in the theoretical analysis. The AS gradient follows distance-dependent connectivity decay. For a given node $i$, its connectivity weights, $W_{FC}(i,:)$, systematically weaken with increasing gradient distance to other nodes, $|\psi_{AS}(i) - \psi_{AS}(:)|$.

This node-specific relationship can be approximated as

$$W_{FC}(i,:) \approx \alpha_i |\psi_{AS}(i) - \psi_{AS}(:)| + \beta_i.$$

Critically, the decay rate $\alpha_i$ is steeper near the gradient extremes reflecting enhanced local specialisation in connectivity proximal to sensory and associative poles, thereby anchoring a global bimodular network structure (Fig 2A, middle panels).

To explicitly quantify how $\psi_{AS}$ spread reflects network specialisation in the connectivity space, linear models were fitted to the rows of the reordered and z-scored connectivity matrices (Fig 2B). Each row, denoted as $r_i = \widetilde{W}_{FC}(i,:)$, was ordered by increasing gradient distance, forming sorted connectivity vectors modelled as:

$$r_i^{(sorted)} \approx \alpha_i x + \beta_i.$$

Here, $x$ represents normalised positions within [0,1], and the slope $\alpha_i$ captures the normalised rate of weight change along gradient positions. Larger magnitudes of $\alpha_i$ at the gradient extremes indicate greater local-distant contrast. At the network level, the vector of transition rates $\alpha = [\alpha_1, \alpha_2, ...]$ was further summarised with a quadratic function

$$\alpha \approx \mathrm{A}x^2 + \mathrm{B}x + \Gamma,$$

where the magnitude of quadratic coefficient |A| provides a global index of weight transition sharpness along $\psi_{AS}$. Networks with larger eigenmode spread (*range*$_{AS}$) displayed greater functional specialisation between sensory and associative modules, characterised by sharper weight transitions along $\psi_{AS}$. In contrast, networks with compressed gradients exhibited more homogeneous connectivity with diminished global relativity (Fig 2B & 2C). Correspondingly, |A| strongly correlated with *range*$_{AS}$ (Fig 2D). This correlation remained robust (r = .77—.80) across varying connectivity thresholds (retaining the top 50—80% of connections per node).

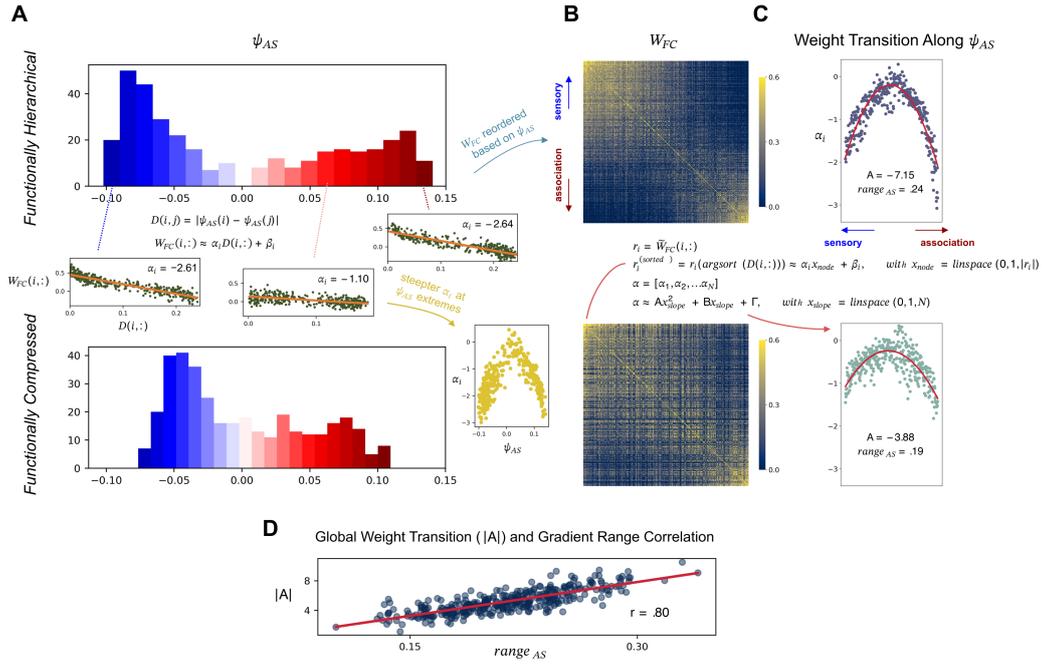

**Fig 2 Gradient spread and weight transition. (A)** The association-sensory gradient ($\psi_{AS}$) distributions for a functionally hierarchical, high-range network (top) and functionally collapsed, low-range network (bottom). Connectivity weights between a given node and others decline with increasing gradient distance (middle). Nodes closer to sensory or associative poles exhibit steeper connectivity decay, indicating greater local specialisation. **(B)** Connectivity matrices ($W_{FC}$) reordered based on regional positions on the $\psi_{AS}$ gradient for a high-range (top) and low-range network (bottom). After z-scoring at the matrix level, each connectivity vector (row) was sorted by gradient distance and fitted using a linear model, resulting in a vector of slope values $\alpha = [\alpha_1, \alpha_2, \dots \alpha_N]$ quantifying connectivity transitions (middle). **(C)** The slope vector ($\alpha$) was further modelled by a quadratic function with normalised gradient positions as the independent variable. The magnitude of the quadratic coefficient (|A|) quantifies the extent to which nodes at gradient extremes exhibit pronounced local-distant contrast, highlighting functionally differentiated, hierarchical organisation. **(D)** Strong positive alignment between the global weight transition coefficient (|A|) and AS gradient range ($range_{AS}$; 60% of the strongest connections retained per node for linear models).

Neural timescale estimation and spatial-temporal convergence

In this section, we investigated the intrinsic relationship between $\psi_{AS}$ and neural timescale without imposing a rigid sensory-association dichotomy. To do so, we first evaluated timescale estimation via simulation comparing a maximum likelihood estimate (MLE) framework based on a generative Ornstein-Uhlenbeck (OU) process to a typical direct exponential fitting of autocorrelation functions (ACF), which risks systematic bias under finite time series length (see Methods: Neural timescale estimation)[44–46].

To validate the OU-MLE approach, we simulated univariate OU time series with a ground-truth timescale, $\tau$, of 3s, 5s, and 10s spanning preliminary empirical estimates from random participants. We varied the sampling interval $\Delta t$ from 1s to 5s in 0.5s increments and the number of time points from 100 to 500 in increments of 50 covering typical fMRI settings. For each parameter combination, we generated 100 random time series instances and fitted them using both methods.

The simulation confirmed the superiority of OU-MLE over direct exponential fitting (Fig 3A), as also seen by Strey (2019)[45]. Under a true $\tau$ of 3s, 5s, and 10s, OU-MLE estimates yielded means $\hat{\tau}$ of 3.02s (mean absolute error mae = .41, variance var = .29), 5.02s (mae = .70, var = .86), and 10.04s (mae = 1.70, var = 5.31), respectively. The exponential fit estimated means $\hat{\tau}$ of 2.93 (mae = .51, var = .47), 4.80s (mae = .99, var = 1.64), and 9.16s (mae = 2.53, var = 9.55). The exponential fit exhibited $\tau$ underestimation, particularly at low $\Delta t$ (1-3s) and short $T$ (100-300 frames), conditions that are especially relevant for empirical fMRI samples. In contrast, the OU-MLE approach remained more robust with less estimation error and variance across conditions.

Having validated OU-MLE, we next examined whether the transition along $\psi_{AS}$ aligns more closely with a simple linear relationship to $\tau$, i.e., lengthening of timescale along the sensory-to-association gradient, or whether continuous specialisation toward the gradient extremes yields a hierarchical temporal structure within both sensory and associative systems. Using the OU-MLE derived $\tau_{OU}$, we compared linear vs quadratic models of $\tau_{OU}$-$\psi_{AS}$ (Min-Max normalised). A paired-sample t-test of Akaike Information Criteria (AIC) showed the quadratic model fit significantly better than the linear model (mean $\Delta$AIC = $-17.8$, p < .001). Under the quadratic model, healthy controls exhibited significantly larger quadratic coefficients $\beta_2$ (mean $\beta_2$ = .50 vs .29) after adjusting for age, gender, site, and TR (Fig 3B; $\beta$ = .16, $p$ < .001, 95% CI = [.07, .25]). This suggests a more pronounced U-shaped $\tau_{OU}$-$\psi_{AS}$ relationship for controls relative to curve obtained for schizophrenia patients indicating diminished relative slowness in the dynamics of $\psi_{AS}$ at extremes.

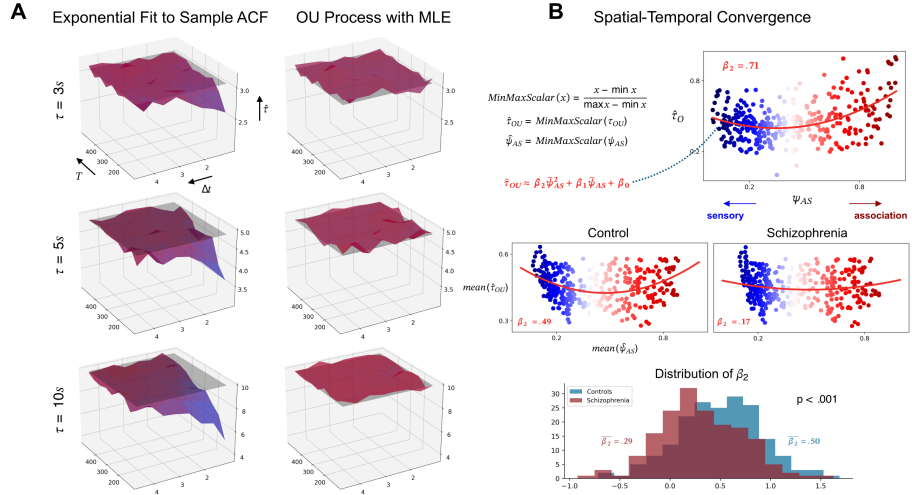

**Fig 3 Neural timescale estimation using an Ornstein-Uhlenbeck process (OU) with maximum likelihood (MLE). (A)** The timescale estimation with ground truth $\tau$ of 3s, 5s, and 10s using direct-exponential (left) and OU-MLE (right); for direct-exponential, the maximal time lag $l_{max}$ was set to half of the sample length. Biased $\hat{\tau}$ with conventional direct-exponential approach was particularly pronounced at low $\Delta t$ (1-3s) and short $T$ (100-300 frames). **(B)** The intrinsic relationship between within-subject Min-Max normalised $\psi_{AS}$ and $\tau_{OU}$ modelled with a quadratic function in an example participant (top), and group averaged representations for both groups (middle). The distribution of quadratic coefficient $\beta_2$ and group comparison via multiple regression revealed a significantly dampened $\tau_{OU}$-$\psi_{AS}$ relationship in schizophrenia.

*Theoretical*

In our theoretical studies, we asked how differences in the spread of the association-sensory (AS) gradient ($range_{AS}$) influence network computation. First, we converted each subject's empirical gradient $\psi_{AS}$ (Fig 4A) into an AS connectivity matrix $W_{AS}$ (Fig 4B) with a generative model based on the gradient-connectivity geometry relationship uncovered earlier. We then regularised a separate recurrent neural network (RNN) per participant, where its recurrent connectivity distributed $W_{AS}$, its inputs channelled into sensory units and predictions read out from association units, all while the network learned working memory (WM) tasks. Once trained, these $\psi_{AS}$-regularised RNNs allowed us to test how gradient spread alters WM learning and the stability of the underlying dynamics.

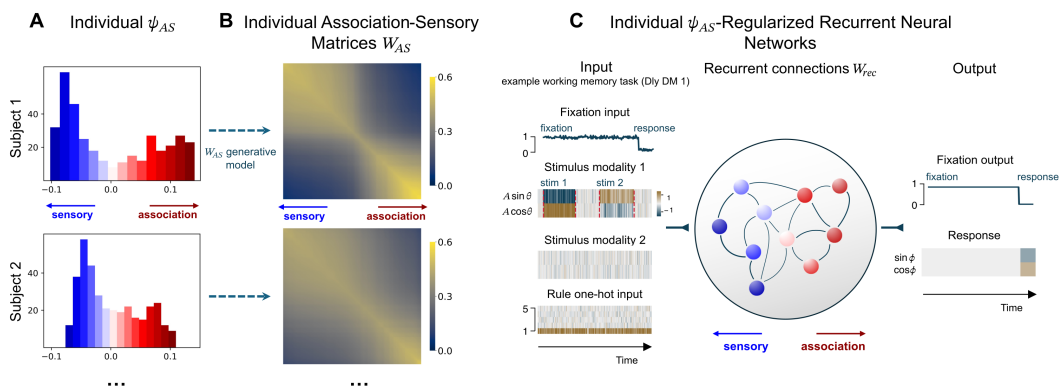

**Fig 4 Pipeline for the theoretical studies. (A)** Subject-specific gradients $\psi_{AS}$. **(B)** Subject-specific AS matrices $W_{AS}$ derived from $\psi_{AS}$ with the generative model (Methods: Generative model of association-sensory matrix). **(C)** For each $\psi_{AS}$-$W_{AS}$ pair, we train a separate RNN. Five WM variants require the network to remember two sequential stimuli and respond in the direction of the stronger one (i.e., higher-amplitude; Methods: Recurrent neural network, working memory tasks, and regularisations). Inputs (noisy; 10 dim): fixation (1), stimulus modality 1 (sin θ, cos θ × amplitude; 2), stimulus modality 2 (2), rule one-hot (5). Depending on the task variant, either one or both modalities are present. Output (3 dim): fixation scalar and [sin ϕ, cos ϕ] of the chosen stimulus.

Generative model of association-sensory weight matrix from $\psi_{AS}$ embeddings

As illustrated in the overview in Fig 4, we first construct each participant's AS connectivity matrix entirely from their empirical gradient $\psi_{AS}$, excluding all other spectral components. We require a robust correspondence between the empirical $\psi_{AS}$ and the principal eigenmode $\psi_{AS}'$ extracted from $W_{AS}$. The resulting $W_{AS}$ then serves as the recurrent-weight constraint in RNN training (Methods: Recurrent neural network, working memory tasks, and regularisations). Because we z-score $W_{AS}$ for RNN regularisation, we consistently extract $\psi_{AS}'$ from the same standardised $W_{AS}$ to maintain methodological alignment.

We tested three distinct approaches for simulating $W_{AS}$ (see Methods: Generative model of association-sensory matrix; Fig 4A):

(1) A purely distance-based locality matrix $W_L$, modelling connectivity decay with gradient separation;
(2) A naïve outer product $\psi_{AS}\psi_{AS}^T$; and
(3) The integrated form $W_L \odot W_G$ that combines local distance-decay and hierarchical scaling thereby enhancing greater local-distant contrast at specialised gradient extremes.

Our primary metrics were the correlation between empirical and recovered gradient ranges ($range_{AS}$ vs $range_{AS}'$), assessing cross-subject gradient spread fidelity; and the mean $\psi_{AS}$-$\psi_{AS}'$ correlations, assessing within-subject preservation of nodal hierarchical orders. The combined $W_L \odot W_G$ yielded correlations of .76 and .92, respectively, surpassing $W_L$ (.52 and .89) and the outer product (.10 and .86). It correlated more strongly with the original connectivity matrix $W_{FC}$ (.43), explaining .20 of the variance compared to .18 or .16 from the other methods.

Despite the robust correlations, the deterministic function of $W_{AS}$ did not sufficiently capture the intended bimodality, instead producing a disproportionally large intermediate grouping between the sensory and association poles (Fig 5B top & bottom). This partial clustering corresponded to sharp connectivity similarity and weight transitions along $\psi_{AS}'$ (Fig 5B middle). Because the density threshold (i.e., 50 connections per node) used in the empirical gradient computation was consistently applied to the deterministic $W_{AS}$, the smoothly decaying function caused specialised nodes to strongly and exclusively interact only their most local connections while aggregating much of the network in the middle of the distribution. A controlled variability would therefore help preserve a more continuous, global similarity pattern.

Indeed, injecting a low magnitude of Gaussian noise mitigated these abrupt boundaries and reinforced the alignment with empirical $\psi_{AS}$ (Fig 5C). Averaged over 100 random realisations, the $range_{AS}$-$range_{AS}'$ and mean $\psi_{AS}$-$\psi_{AS}'$ correlations were improved to .94 (std = .001) and .98 (std < .001), respectively. Of note, the alignment of $\psi_{AS}'$ from z-scored and original $W_{AS}$ under both metrics was $\approx 1$. Thus, coupling local-global generation with controlled stochasticity enhances the biological plausibility while achieving faithful recovery of the AS eigenmode within $W_{AS}$.

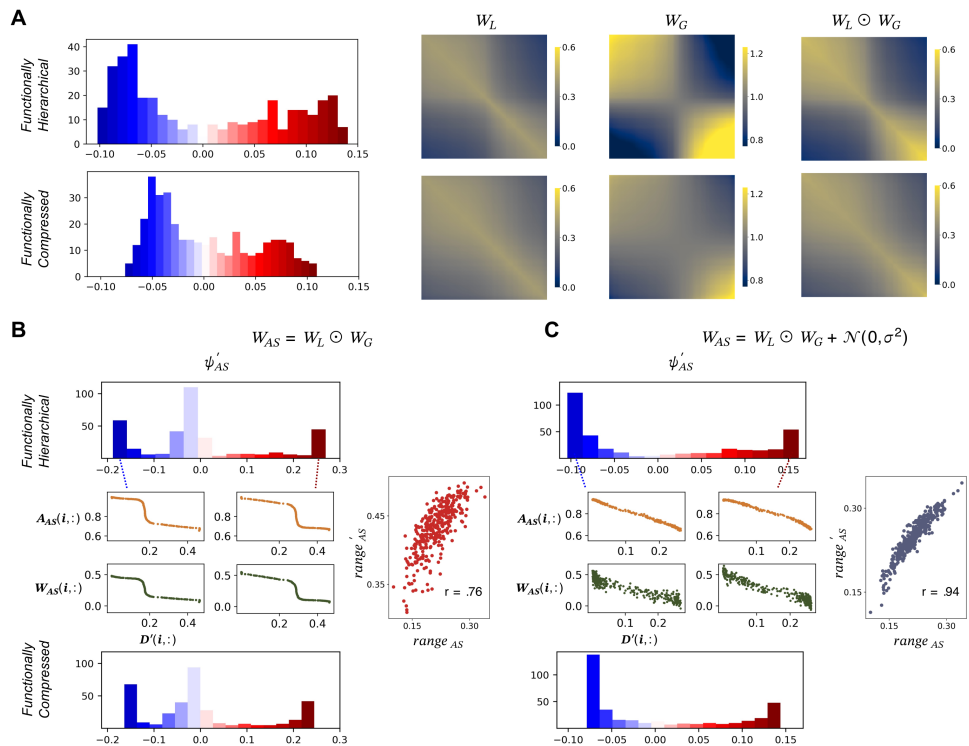

**Fig 5 Association-sensory matrix ($W_{AS}$) generation. (A)** Distributions of empirical association-sensory gradient values color-coded from sensory (blue) to association (red). On the right are three different matrix constructions, (1) a locality matrix $W_L$ modelling connectivity decay with gradient distance using a node-invariant decay rate; (2) a global matrix $W_G$ based on scaled outer product of the gradient; and (3) their elementwise product $W_L \odot W_G$, for a functionally hierarchical (broader $range_{AS}$; top) and collapsed (reduced $range_{AS}$; bottom) networks. **(B)** The principal gradient derived from $W_{AS} = W_L \odot W_G$. Across subjects, the recovered gradient range ($range_{AS}'$) correlated at .76 with the empirical $range_{AS}$. The deterministic form showed a large intermediate grouping along the transition from sensory (blue) to association (red) poles (top & bottom), with abrupt connectivity similarity and weight transitions (middle). **(C)** The principal gradient computed when variability was introduced—$W_{AS} = W_L \odot W_G + \mathcal{N}(0, \sigma^2)$, with $range_{AS}$-$range_{AS}'$ correlation improved to .94 and smooth transitioning between sensory and association systems.

Recurrent neural network and working memory task learning dynamics

Having generated $W_{AS}$ from $\psi_{AS}$, we next assessed the computational and dynamical consequences of varying $\psi_{AS}$ spread. To this end, we trained one continuous-time RNN per subject-specific $W_{AS}$-$\psi_{AS}$ pair on a family of parametric working memory (WM) tasks and constrained its recurrent weights toward the corresponding $W_{AS}$ (Fig 4). We implemented an RNN architecture and learning protocol similar to previously studied multitask frameworks (detailed in Methods: Recurrent neural network, working memory tasks, and regularisations)[33,34]. Each WM trial presents two brief stimuli (circular variables with both angle and amplitude), separated by delays, and the network is required to report the stronger stimulus[33,34,47].

Regularisation had three layers (Fig 6A):

(1) Recurrent connectivity constraint—a mean-squared error loss pulling the z-scored absolute recurrent connectivity $|W_{rec}|$ toward the z-scored $W_{AS}$, preserving sign flexibility while enforcing AS geometry;

(2) Input loading—an L1 term directs lower-order (sensory) units to receive external inputs; and

(3) Output routing—a complementary L1 term encourages higher-order (associative) units to drive decisions.

To reduce computational load, our RNN training was based on a single site (Hartford, selected for the largest balanced cohorts; $N_{HC} = 45$, $N_{SZ} = 47$). We trained with five independent random initialisations (seeds) per subject (92 × 5 networks) for 20,000 steps, smoothed learning curves with a 500-step moving average to reveal denoised trends, and tracked Spearman correlations between $range_{AS}$ and both task loss and weight-regularisation loss throughout training.

During the initial phase, total losses did not exhibit marked divergence across networks with varying $range_{AS}$. However, after ~4,000 steps, those constrained by a broader AS gradient descended more steeply, converging onto lower plateaus (see color-coded curves in Fig 6B). Meanwhile, $W_{rec}$ regularisation loss $L_{weights}$ initially reached a minimum, followed by an oscillatory phase, diverging and stabilising according to $range_{AS}$ as WM learning intensifies. Correspondingly, WM performance displayed an emergent separation by $range_{AS}$, with broader eigenmodes yielding steeper performance gains. Sigmoid fits to the WM trajectories showed a positive correlation between steepness and $range_{AS}$ (Fig 6C; mean Spearman's r = .62 ± .05 across seeds). The WM trajectories of task variants were similar and plateaued at comparable levels (Supplementary Fig 2). Overall, this pattern suggests a two-stage learning process—networks first aligned with the AS gradient constraints, then adaptively reconciled that alignment to meet task demands.

Networks endowed with a broader gradient more readily accommodated both regularisation and functional demands, more efficiently adopting AS-like structure while simultaneously advancing toward stronger WM performance. This was evidenced by systematic decays in the correlation between $range_{AS}$ with both WM ($L_{task}$) and regularisation ($L_{weights}$) losses, which stabilised by ∼10,000 steps at $-.61 \pm .04$ and $-.65 \pm .06$, respectively (Fig 6C). Furthermore, similar correlation trajectories were observed using the gradient spread computed on trained recurrent matrices $W_{rec}$, with $L_{task}$-$range_{AS}'$ and $L_{weights}$-$range_{AS}'$ correlations settling at $-.69 \pm .04$ and $-.72 \pm .05$, respectively, reinforcing the WM task advantage conferred by broader gradients.

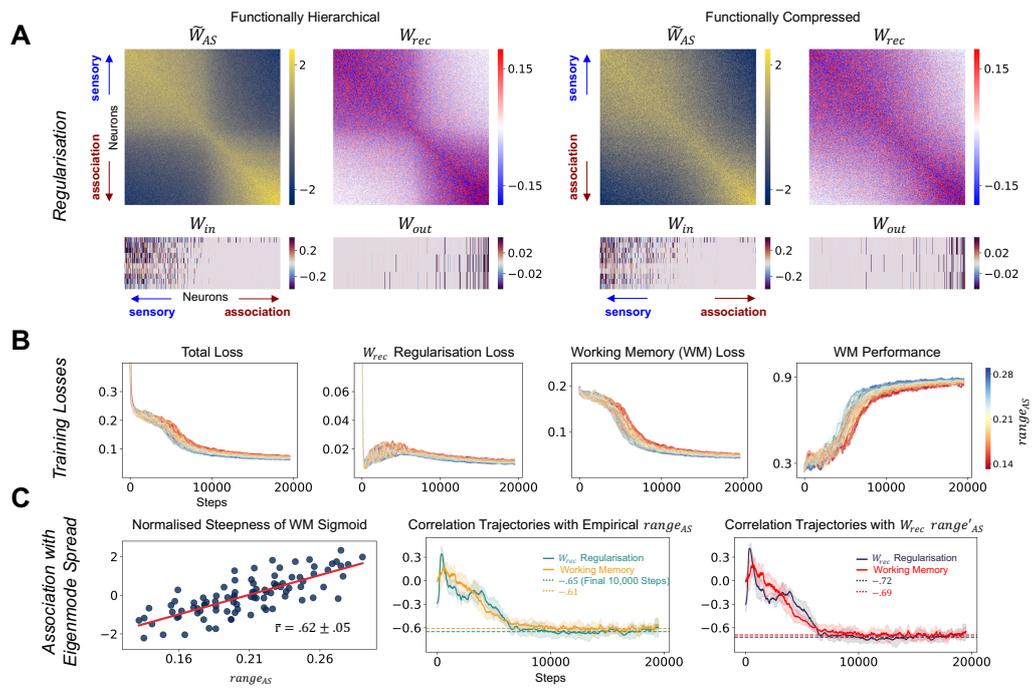

**Fig 6 Learning dynamics of association-sensory gradient ($\psi_{AS}$)—regularised recurrent neural networks. (A)** Example z-scored AS matrices $W_{AS}$ derived from $\psi_{AS}$ (see Fig 2 and 4), shown for functionally hierarchical (left) and functionally compressed (right) networks, along with the corresponding regularised recurrent ($W_{rec}$), input ($W_{in}$), and output ($W_{out}$) weight matrices post-training. The $W_{rec}$ regularisation enforces the AS organisational pattern in $W_{AS}$, while the $W_{in}$ and $W_{out}$ regularisations facilitate input loading and prediction routing based on nodal specialisations implied by $\psi_{AS}$ (Methods: Recurrent neural network, working memory tasks, and regularisations). **(B)** Training curves for total loss, $W_{rec}$ regularisation loss, working memory (WM) task loss, and WM performance, each smoothed over 500 steps. Colour code reflects the $\psi_{AS}$ used to generate $W_{AS}$, with bluer lines corresponding to broader AS gradient ranges. **(C)** Correlation between the empirical $range_{AS}$ used for $W_{AS}$ generation and the steepness parameter $k$ of sigmoid fit to the WM performance, $f(x) = L/(1 + e^{-k(x-x_0)})$ (left), correlation trajectories between $range_{AS}$ and both $W_{rec}$ regularisation loss (green) and working memory loss (yellow; middle), as well as the trajectories using $range_{AS}'$ computed from trained $W_{rec}$ (right).

Stability during delay epochs via linearisation around fixed points

After showing that a broader $\psi_{AS}$ accelerates and deepens WM task learning, we sought deeper mechanistic insight by assessing the stability of the network's mnemonic states. We therefore performed fixed point (more precisely, slow points) analysis during the two post-stimulus delay epochs (Memory 1 and Memory 2; Fig 7A), where the network settles into population states that maintain stimulus information once input ceases. Candidate states $h$ were sampled from each trained RNN optimised toward minimal update $\|h - F(h, u)\|$; those with small residual motion represent locations where network activity effectively stabilises under constant input $u$ (see Methods: Fixed point linearisation analysis)[34,41].

To focus on functionally relevant slow points, we ranked the ~1,000 candidates per epoch by this residual motion, i.e., the energy, and retained the 100 lowest-energy points ($\approx$ top 10%). Lower energy denotes the state is closer to an exact fixed point; the filtering hence sharpens stability estimates. For each retained point we computed the Jacobian $J(h^*)$; its largest eigenvalue magnitude $|\lambda|$ indexes the most unstable direction. Averaging these maxima served a network-level stability metric.

Across all task rules and seeds, the average energy cutoffs (for filtering) were $1.54 \times 10^{-6}$ for Memory1 and $1.33 \times 10^{-6}$ for Memory 2, with lower cutoffs seen in networks with higher $range_{AS}$ (Memory 1: Spearman's r = −.26, p = .02; Memory 2: r = −.23, p = .03). Correspondingly, the mean energy of all slow points showed significant negative correlation with $range_{AS}$ (Memory 1: r = −.48, p < .001; Memory 2: r = −.44, p < .001; Fig 7B), as well as with $range_{AS}'$ of trained RNNs ($W_{rec}$; Memory 1: r = −.50, p < .001; Memory 2: r = −.45, p < .001). These observations suggest that systems with broader gradient spread converge to lower-energy slow points, with their neural states functionally closer to true fixed points.

Furthermore, analysis of local stability via the maximum eigenvalue magnitudes revealed that networks with higher $range_{AS}$ showed smaller mean maximum $|\lambda|$ at both memory epochs (Memory 1: Spearman's r = −.42, p < .001; Memory 2: r = −.38, p < .001; Fig 7B), and similarly for $range_{AS}'$ (Memory 1: r = −.42, p < .001; Memory 2: r = −.37, p < .001). Recall that in the unstable regime $|\lambda| > 1$, approaching unity corresponds to dynamics along the corresponding eigenvector at that state which are closer to marginal stability. These results indicate that even in the most unstable directions, networks with broader gradient spread achieve greater local asymptotic stability of memory dynamics. Taken together, our findings establish a link between greater association-sensory eigenmode spread and more dynamically stable maintenance of states during memory delay.

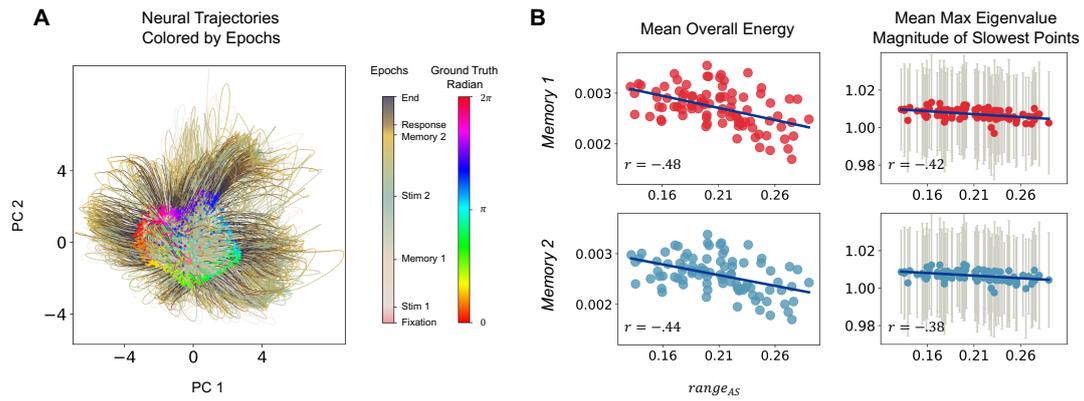

**Fig 7 Greater association-sensory gradient ($\psi_{AS}$) spread is associated with more stable network states during memory delay.**
**(A)** Low-dimensional neural trajectories for a trained RNN performing Delayed Decision-Making (Dly DM) 1 task. Each trajectory (of a trial) is color-coded by epoch, and trajectory endpoints are color-coded by the target stimulus radian. Neural states for slow point optimisation were sampled from Memory epochs. **(B)** Mean energy of all slow points and mean maximum $|\lambda|$ of the filtered slow points (with error bars denoting the standard deviations) both showed negative correlations with the range of $\psi_{AS}$ used to regularise the RNN.

**Discussion**

In this study, we focused on the most overarching organisational axis of functional brain systems, the association-sensory (AS) gradient. Rather than discrete functional modules, the AS gradient encodes a continuous transition of connectivity patterns from unimodal to transmodal regions. Our experimental analyses revealed that schizophrenia is characterised by a de-differentiation of AS gradient and a dampened relationship between gradient extremes and neural timescales, suggesting that dysconnectivity manifests as a contraction of the brain's hierarchical organisation in both spatial and temporal domains. Meanwhile, in our theoretical branch, we show that more diffused and well-differentiated AS systems support more adaptive learning of a canonical working memory (WM) computation and, critically, maintain network states more stably during memory delays.

The compression of the principal AS gradient in schizophrenia aligns with prior evidence, indicating a functional contraction, i.e., less differentiated specialised systems, along the brain's most global organisational dimension[14,15,48]. Discrete dysconnectivity studies corroborate this pattern, showing diminished within-network coherence in both sensory and higher-order systems, as well as less efficient multistep connectivity propagation along the AS hierarchy in schizophrenia[14,49–51]. As demonstrated empirically and via $W_{AS}$ generative model, broader $\psi_{AS}$ spread reflects a more pronounced hierarchical arrangement, with steeper connectivity transitions and stronger local connectivity among similarly specialised nodes, and sparser connectivity among distant gradient extremes. In contrast, a compressed gradient corresponds to an attenuated local-distant contrast, resulting in flatter, less differentiated architecture that diverges from the sparsely inter-connected specialised compositions believed to foster adaptive, concurrent, and locally segregated processes[52,53]. Correspondingly, gradient range was positively associated with cognition in the entire sample and negatively associated with positive symptoms in the schizophrenia cohort. Beyond connectivity, the AS transition is evident in latent dimensions of neural timeseries, in dynamical property decomposition, and in the formation of low-energy attractor states that exert a "gravitational pull" on brain activity configurations[54–56]. Consequently, this reduced specialisation in connectivity space may critically undermine stability of neural dynamics and computations, a hypothesis we explored via timescale analyses and AS gradient-regularised RNNs.

Our analyses revealed that neural timescales tend to lengthen with greater functional specialisation. Dynamic profiles of the brain are not merely local phenomena; for instance, regions with stronger power in lower frequencies, akin to low-pass filtering, and slower timescales display higher functional connectivity (FC), both globally and within unimodal and transmodal systems[17,57]. In parallel, regional dynamic profiles are more similar within rather than between functional modules, and longer timescale has been specifically linked to within-community integration, where hubs at the core of specialised systems can facilitate segregated information processing[26].

In contrast, faster, more flexible dynamics, indicative of enhanced sensitivity toward instantaneous signals, may characterise connector hubs that bridge distinct functional subsystems, as suggested by their shifting synchronisation and modular allegiance over time[58]. These hubs dynamically coordinate inter-module communication without overloading each specialised systems, thereby preserving local autonomy[59]. Functionally, such transitional systems act as saddle points, routing transient signals onto the slower, more specialised regions that then serve as stable attractors for sustained encoding[41].

We did not explicitly model the unimodal-transmodal dichotomy, where prior work has reported overall faster dynamics in unimodal systems hosting perceptual processes and slower transmodal dynamics supporting integrative functions[17,22,23]. Instead, our analysis offers a complementary perspective on spatial-temporal convergence, whereby timescales along $\psi_{AS}$ appear to mirror the quadratic organisation of weight transition rates, such that specialised anchors within each subsystem display slower neural dynamics. Nevertheless, although the quadratic model outperformed a simple linear fit, the mapping remains individually heterogenous (e.g., coefficient strength) and thus requires further validation.

The AS gradient-timescale mapping in schizophrenia showed a less pronounced hierarchical organisation of neural dynamics, with diminished relative slowness at the gradient extremes. Previous findings suggested globally shortened timescales, symptom-specific hierarchical disruptions, and instability in dynamic FC that indicate more rapidly fluctuating neural synchronisations in schizophrenia[24,60–62]. The observed reduction in slowness at specialised poles may further reflect destabilised cortical attractor states that are prone to noise-provoked random transitions, undermining sustained information encoding[63,64]. Having linked spatial gradients to temporal dynamics empirically, we next interpret how these observations resonate with our RNN results.

In our $\psi_{AS}$-regularised RNNs, networks with a higher gradient range achieved more efficient learning and lower task loss, while maintaining lower regularisation loss against the AS connectivity constraint. Such well-differentiated functional systems may be less susceptible to computational interference. This synergy between the high-range regularisation and the task demand is consistent with the simultaneous yet autonomous computations performed by specialised neural systems, coordinated via connector nodes in the brain[59,65,66].

Using our $W_{AS}$ generative model, we integrated stereotypical features of brain network topology—functional modularity, sparsity between specialised systems, small-worldness—yet sustained a continuous approach without imposing rigid modules. These topological properties are theorised to arise from the joint optimisations of metabolic costs (i.e., network development and maintenance) and information-processing demands[67,68]. Theoretical work suggests that balancing local cohesion and global diffusion fosters effective signal spreading and computations[69].

On the other hand, functional demands actively sculpt network organisation, as evidenced by RNN hidden units that self-organise into compositional, specialised functional clusters, thereby support cognitive flexibility[33]. Similarly, explicitly enforcing biophysical wiring constraints in RNNs can generate structural motifs and functional clustering reminiscent of brain networks when optimised for inference tasks[68]. By incorporating a gradient-constrained connectivity, our framework echoes with these findings, suggesting that a more diffuse functional organisation, featuring steeper local-distant contrast, may achieve an economical balance between computational efficiency and connectivity cost, a principle seemingly shared between biological brains and artificial RNNs.

Moreover, greater $\psi_{AS}$ range was associated with more stable, slowly evolving neural states during memory delay periods. This corroborates studies showing that artificial networks with modular, hierarchical functional communities exhibit persistent activities and stable, scalable activation spread, unlike random networks[70–72]. In the brain, stable maintenance of whole-brain activity patterns correlates with greater WM performances, whereas in schizophrenia, brain state stability is reduced, harder to control, and more vulnerable to perturbations[73–75]. This pattern aligns with dynamical system models of WM deficits in schizophrenia, which posits a flattened attractor landscape and corresponding unstable, lower-capacity memories susceptible to distractibility[27,76–78]. Our RNN simulations align with this view and, critically, offer a testable hypothesis, where task-based designs that jointly model gradient geometry, brain state stability, and trial-wise task performances can determine whether AS contraction mechanistically undermine state maintenance and cognition. Taken together, we present two complementary perspectives on the instability of neural dynamics in schizophrenia; one from experimental evidence of timescale flattening, and another from theoretical demonstration of more unstable fixed point dynamics exhibited by low-range RNNs.

All FC used for gradient extraction and computational models was derived from resting-state data. By imposing these FC-based constraints, we capitalise a direct proxy for interareal communication, as the activity flow principal—modelling a region's activation as the sum of inputs weighted by empirical resting-state FC—predicted task-state activations[55,79,80]. Critically, the brain's resting-state network architecture fundamentally shapes its task-based organisation and maintains a full repertoire of interacting networks even at "rest"[59,81,82]. As a regulariser, the resting architecture also highly aligns with multitask FC, reinforcing the notion of a stable functional scaffold at rest[83]. Although task-fMRI typically reveals necessary network integration when tasks are not fully automated, training has been shown to drive more pronounced functional segregation of specialised systems, thereby boosting execution autonomy and neural efficiency[84,85].

In this context, our study focuses on the schizophrenia-implicated AS gradient, examining its role within this standard functional architecture and its computational consequences. Notably, task-state AS gradient spread has been shown to be predictive of cognition, motivating further validation of our observations with task-based gradients in schizophrenia and optimisation of gradient-based regularisation strategies[86]. While our current RNNs employ regularisation based solely on the AS gradient, future studies should consider incorporating additional gradients, such as the cross-sensory gradient, to explore how multiple connectivity dimensions interact to shape computations.

To conclude, through a continuous-network perspective, we investigated dysconnectivity in schizophrenia along the brain's principal AS gradient and its dynamical and computational consequences. Empirically, we showed that schizophrenia displays gradient compression indicative of reduced hierarchical functional specialisation, paralleled by diminished temporal differentiation. Theoretically, embedding these empirical gradient structures into RNNs showed that AS system compression destabilises fixed point dynamics critical for noise-robust computational stability. This integrated experimental-theoretical framework not only advances connectivity-to-computation insights into schizophrenia, but also establishes a foundational approach for future neuroAI research. In particular, it sets a precedent for designing artificial systems informed by spectral properties of brain connectivity, enabling wider empirically-grounded theoretical explorations of neural computation.

**Materials and methods**

*Experimental*

Functional MRI

The resting-state fMRI data from BSNIP consortium consisted of 186 healthy controls and 169 schizophrenia patients; identical diagnostic and recruitment approaches were applied across sites, and all subjects underwent a 5-minute resting-state scan on a 3-T scanner [42,43]. The fMRI images were preprocessed by a prior pipeline that included slice-time correction, rigid-body head motion correction, co-registration to the T1-weighted anatomic volume, transformation to MNI152 standard space, wavelet despiking of motion artefacts, regression of 12 motion parameters, and spatial smoothing at 6 mm FWHM[87,88]. Cortical parcellation was conducted using the Glasser atlas with 360 regions[89]. Time series were then bandpass filtered using wavelet scales 2 and 3 (covering 0.028-0.167Hz across the dataset), leveraging wavelets' simultaneous time-frequency localisation to mitigate the influence of long-memory processes[90–94]. Functional connectivity (FC) matrices were derived as the pairwise Pearson correlation between time series from all region pairs.

Spatial gradients via diffusion map

Pairwise affinity $A$ from each sparsified FC matrix was computed using normalised angle kernel. To extract principal eigenmodes, we first constructed a kernel $K = D^{-\alpha} A D^{-\alpha}$, where $D$ is the degree matrix $D_{ii} = \sum_j A_{ij}$, and $\alpha = 0.5$, approximating Fokker-Planck diffusion. A Markov diffusion operator was then computed as $P = (\sum_j K_{ij})^{-1} K$, whose eigenvectors were scaled with multiscale eigenvalue aggregates (i.e., $\sum_{t=1}^{\infty} \lambda_k^t = \lambda_k / (1 - \lambda_k)$), producing set $\{\psi_k\}$ that defines functional gradients[95–97]. This process utilised the BrainSpace toolbox[97].

A quantitative measure of network hierarchy encoded by each gradient is given by its range, $range(\psi_k) = \max(\psi_k) - \min(\psi_k)$, where a larger spread indicates steeper transitions among network elements along that diffusion axis. Individual embeddings were aligned to the group-level representation using orthogonal Procrustes transformations, which preserve inter-point Euclidean relations through rigid rotations without scaling, ensuring dimensional consistency across all embeddings.

Connectivity matrices were sparsified at three density thresholds, retaining the top 40, 50, and 60 edges per node (approx. 11%, 14%, and 17% of the strongest connections, respectively). These thresholds align with thresholds used in previous studies and serve to balance the signal-to-noise ratio. After confirming the robustness of gradient patterns across all thresholds, we adopted the intermediate density of 50 edges per node (14%) for subsequent analyses.

Neural timescale estimation

We define the timescale $\tau$ of a neural time series $\{X_t\}$ by the exponential decay rate of its autocorrelation function (ACF), $ACF(s) = \mathbb{E}[X_t X_{t+s}]/\mathbb{E}[X_t^2]$. A typical estimation approach is to fit an exponential directly to empirical ACF values. However, for finite sample sizes, such estimates can exhibit systematic biases that depend on both the sample series length and the underlying ACF, see Zeraati et al. (2022)[46].

To mitigate these biases, we model $\{X_t\}$ as an Ornstein-Uhlenbeck (OU) process, a Gauss-Markov process with mean reversion and a closed-form ACF. It satisfies the stochastic differential equation

$$dX_t = -\frac{1}{\tau} X_t dt + \sigma dW_t,$$

where the deterministic term $-\frac{1}{\tau} X_t dt$ (with characteristic timescale $\tau > 0$) pulls $X_t$ toward its long-term mean (assumed zero in mean-centred data), while the stochastic term $\sigma dW_t$ (with noise amplitude $\sigma > 0$ and $W_t$ as a standard Wiener process) injects noise. In stationarity, its autocorrelation decays exponentially as

$$ACF(s) = e^{-\frac{|s|}{\tau}}.$$

Discretising at sampling interval $\Delta t$ yields a Markov sequence where

$$\mathbb{E}[X_{t+1} \mid X_t] = X_t e^{-\frac{\Delta t}{\tau}}, \ Var[X_{t+1} \mid X_t] = D\tau(1 - e^{-\frac{2\Delta t}{\tau}}),$$

with $D = \sigma^2/2$. By Markov property, the joint likelihood of the entire observed sequence is the product of conditional Gaussian densities

$$L(\tau, D \mid X_0, X_1, \ldots, X_T) = \prod_{t=0}^{T-1} \frac{1}{\sqrt{2\pi Var[X_{t+1} \mid X_t]}} \exp\left[-\frac{(X_{t+1} - \mathbb{E}[X_{t+1} \mid X_t])^2}{2 Var[X_{t+1} \mid X_t]}\right].$$

Taking the negative log-likelihood (NLL) in terms of parameters $\tau, D$ gives

$$-\log L(\tau, D) = \sum_{t=0}^{T-1} \left[\frac{1}{2}\log\left(2\pi D\tau\left(1 - e^{-\frac{2\Delta t}{\tau}}\right)\right) + \frac{\left(X_{t+1} - X_t e^{-\frac{\Delta t}{\tau}}\right)^2}{2D\tau\left(1 - e^{-\frac{2\Delta t}{\tau}}\right)}\right].$$

Minimising this NLL with respect to $\tau$ and $D$ produces maximum likelihood estimates (MLE) $\hat{\tau}$ and $\hat{D}$.

For comparison, a computationally straightforward approach is to compute the empirical autocorrelations $\hat{\rho}(\ell)$ at each time lag $\ell$ (up to a predefined maximal lag). These $\{\hat{\rho}(\ell)\}$ are then jointly fitted to an exponential decay $e^{-\ell\Delta t/\tau}$ directly with least-squares. However, each $\hat{\rho}(\ell)$ carries estimation variability, in contrast to OU-MLE inferring $\tau$ from a complete probabilistic model of the data. The Gaussian assumption of OU was empirically validated by assessing stationarity of the preprocessed time series (Augmented Dickey-Fuller) and Gaussianity of both the OU-fitted residuals and time series (Shapiro–Wilk) in randomly sampled subjects. Having established more reliable estimates of $\tau$ via OU-MLE, we then examined how these intrinsic timescales align with the spatial gradient.

Spatial-temporal convergence

To assess the relationship between the timescale $\tau$ and association-sensory (AS) gradient $\psi_{AS}$, we normalised $\tau$ and $\psi_{AS}$ using $MinMaxScalar(x) = \frac{x - \min x}{\max x - \min x}$ within each subject to account for cross-subject variability in absolute ranges and ensure comparability of relative patterns. We then compared a linear $\hat{\tau} \approx \beta_1 \hat{\psi}_{AS} + \beta_0$ and quadratic $\hat{\tau} \approx \beta_2 \hat{\psi}_{AS}^2 + \beta_1 \hat{\psi}_{AS} + \beta_0$ model, where the quadratic term $\beta_2$ captures potential curvature indicative of hierarchical temporal organisation. Model selection was guided by paired AIC comparison across subjects. For the selected model, case-control comparison of the coefficient of interest—i.e., $\beta_1$ for linear, $\beta_2$ in quadratic—were tested using multiple regression adjusted for confounding variables.

*Theoretical*

Generative model of association-sensory matrix $W_{AS}$

To investigate how variations in AS gradient $\psi_{AS}$ influence working memory computations with recurrent neural networks (RNNs), we developed a generative model to construct subject-specific AS connectivity matrices $W_{AS}$ using their own empirical $\psi_{AS}$, while holding the model parameters fixed. These matrices, whose principal gradients preserve each subject's nodal hierarchy and gradient range, were then used to regularise RNNs, enabling us to directly assess the effect of gradient spread on task learning and dynamical stability.

Our generative process combines distance-dependent connectivity decay with hierarchical scaling, building on empirical observations from Results: Eigenmode spread and network specialisation. For each subject, we first modelled the observed connectivity decay with gradient separation by defining a locality matrix

$$W_L = \alpha D + \beta, \quad D_{ij} = |\psi_{AS}(i) - \psi_{AS}(j)|.$$

We fit linear decay models at each node on the mean connectivity matrix $\overline{W}_{FC}$ across all subjects then averaged those fits across nodes to obtain the node-invariant parameters $\alpha = -1.35$ and $\beta = -.40$. Although $W_L$ encodes local granularity along each subject's $\psi_{AS}$, it does not emphasise node specialisation, characterised by steeper weight drop-off (higher $|\alpha|$) at sensory or association poles. To address this, we elementwise multiplied $W_L$ by

$$W_G = \tilde{\psi}\tilde{\psi}^T + 1, \quad \tilde{\psi} = \frac{\psi_{AS}}{\gamma}.$$

This positive modulation differentially augments ($W_{G\,ij} > 1$) or dampens ($W_{G\,ij} < 1$) connections based on nodes' hierarchical positions along that subject's gradient, effectively adjusting node-level $\alpha$. The optimal scaling parameter $\gamma$ for gradient recovery was observed within .15—.25, we selected an intermediate $\gamma = .20$ achieving realistic connectivity ranges (average min = .08, max = .57, mean = .32) while ensuring gradient fidelity. The resulting $W_L \odot W_G$ outperformed simpler distance-only or outer-product approaches.

However, this deterministic function produced overly discrete boundaries, we therefore introduced low-intensity Gaussian noise ($\sigma = .05$) to smooth sharp transitions (see Results: Generative model of association-sensory weight matrix from $\psi_{AS}$ embeddings), with $\sigma$ determined by incremental testing (from 0 to .10 in .01 steps) to ensure robust recovery of gradient spread (boosting correlation from .76 to .94) without distorting the overall hierarchy.

Our final generative model was thus

$$W_{AS} = W_L \odot W_G + \mathcal{N}(0, \sigma^2),$$

ensuring strong alignment between the recovered $\psi_{AS}'$ and the empirical $\psi_{AS}$, while enhancing the smoothness of the transition between sensory and association systems.

Recurrent neural network, working memory tasks, and regularisations

We examined Euler-discretised continuous-time RNNs[33,34], following leaky integration update

$$h_{t+1} = (1-\alpha)h_t + \alpha f(W_{rec}h_t + W_{in}u_t + b + \xi_t),$$

$$f(x) = max(0,x).$$

Here, the integration constant $\alpha = \Delta t/\tau$ governs information retention and was set to .20. $W_{in}$ and $W_{rec}$ are the input and recurrent weight matrices of dimensions $N_{in} \times N_{neuron}$ and $N_{neuron} \times N_{neuron}$, loading external inputs $u_t$ and upstream activities to downstream units. The noise term $\xi_t$ comprises $N_{neuron}$ independent Gaussian white noise processes scaled by $.05\sqrt{2\alpha^{-1}} \approx .158$. Output units $z$ were computed via linear readout

$$z_t = W_{out}h_t + b_{out},$$

with $W_{out} \in \mathbb{R}^{N_{neuron} \times N_{out}}$ and bias $b_{out}$.

Working memory tasks were adopted from delay decision making (Dly DM) task family (five variants; Fig 4) in Yang et al.[33] and Driscoll et al.[34]. Briefly, noisy inputs $u$ included a one-dimensional fixation signal, four-dimensional stimulus channels, and five-dimensional rule one-hot vectors. Stimuli were angles θ∈[0,2π), presented in two modalities as [$Asin\theta, Acos\theta$] pairs, with $A$ modulating strength. Trial epochs comprised initial fixation, stimulus 1, memory 1, stimulus 2, memory 2, and response, with each duration uniformly sampled from preset ranges. In "Dly DM 1 & 2" tasks, only one modality was available; in "Ctx Dly DM 1 & 2" tasks, both modalities appeared but only one was attended; and in "Mult Dly DM" tasks, both modalities informed decision. The task objective was to accurately select the stimulus direction of highest intensity, with performance is deemed accurate if the network's final response angle $\phi$ lay within $\pm\pi/5$ of the target. Target output $\hat{z}$ was a fixation component plus [$sin\phi, cos\phi$] encoding the chosen angle. We computed squared error loss between $\hat{z}$ and network output, weighted by a cost mask accentuating post-response errors. Training was conducted simultaneously on all five task variants over 20,000 iterations using minibatches of 128 trials, with each unique batch randomly drawn from one of the variants with equal probability. Networks were optimised with Adam (learning rate = 10^-3), a variant of stochastic gradient descent, to minimise the total loss (see below)[98].

To regularise $W_{rec}$—initialised with random orthogonal initialisation—toward hierarchical patterns resembling the AS constraint $W_{AS}$ generated from $\psi_{AS}$, while minimising sensitivity to raw magnitude, we defined a weight-regularisation loss $L_{weights}$ as the mean-squared error between z-scored $\widetilde{W}_{AS}$ and z-scored, absolute $\widetilde{W}_{rec}$

$$\ell_{weights} = MSE(\widetilde{W}_{AS}, \widetilde{W}_{rec}).$$

Both matrices were mean-centred and scaled to unit variance, ensuring organisational alignment rests on connectivity pattern rather than penalising overall weight scales. Taking $|W_{rec}|$ before standardisation aligns its relative magnitude profile with the functional interactions encoded in $W_{AS}$, without constraining recurrent links to be excitatory or inhibitory, preserving functional flexibility.

To conform each network's input and output structure with its intrinsic nodal AS specialisations, we imposed L1 penalties on $W_{in}$ and $W_{out}$. Specifically, $\psi_{AS}$ was min-max scaled into $\tilde{\psi}_{AS} \in [0,1]$, mapping sensory neurons toward 0 and association to 1. During training, we computed

$$\ell_{in} = \langle \tilde{\psi}_{AS} \odot |W_{in}| \rangle \quad and \quad \ell_{out} = \langle (1 - \tilde{\psi}_{AS}) \odot |W_{out}| \rangle.$$

The network was therefore encouraged to direct inputs primarily into lower-order sensory units while routing predictions through high-order association regions. Our total loss was defined as

$$\ell_{total} = \ell_{task} + \ell_{weights} + \ell_{in} + \ell_{out}.$$

Fixed point linearisation analysis

The dynamics of trained, $\psi_{AS}$-regularised RNNs were examined through identifying fixed points—states $h^*$ with which as initial conditions, the system exhibits minimal motion and satisfies $h^* \approx F(h^*, u)$; $F$ is the update rule and $u$ is an external input vector specifying the task condition[34,41]. Our fixed points included approximately fixed slow points where the system is not stationary but evolves minimally.

Near each slow point, $h^* + \delta h_t$, the state update can be linearised through first-order Taylor expansion

$$h_{t+1} = F(h^* + \delta h_t, u) \approx F(h^*, u) + J(h^*)\delta h_t,$$

Where $J(h^*)$ is the Jacobian matrix evaluated at $h^*$, with

$$J_{ij}(h^*) = \left.\frac{\partial F_i}{\partial h_j}\right|_{h=h^*}.$$

The evolution of sufficiently small perturbations, $\delta h_t$, around a slow point can thus be approximated as

$$\delta h_{t+1} \approx J(h^*)\delta h_t.$$

The local stability, i.e., whether the system converges to or diverges from $h^*$, was assessed through the eigendecomposition of $J(h^*)$. The system contracts along a dimension (eigenvector) if the corresponding eigenvalue magnitude $|\lambda| < 1$ (locally stable), expands if $|\lambda| > 1$ (unstable), and is marginally stable if $|\lambda| \approx 1$. Our analyses focused on post-stimulus delay epochs, where the network's hidden activity holds stimulus information under the constant fixation input.

For each task variant and $\psi_{AS}$-regularised RNN, we generated a batch of trials and sampled 1,000 random neural states $h$ from each of the Memory 1 and Memory 2 epochs. We then optimised these states by gradient descent on the energy function

$$q = \tfrac{1}{2}\|h - F(h, u)\|^2,$$

to identify candidate slow points. From the resulting ~1,000 slow points (per epoch), we sorted by energy, retained the 100 slowest-energy points, and computed at each the maximum eigenvalue magnitude $|\lambda|$ of $J(h^*)$. Finally, we averaged (1) the energies of all slow points and (2) the $|\lambda|$ maxima of the filtered subset across task variants and random seeds, then correlated both metrics with each network's AS gradient range to assess how gradient spread relates to memory dynamics.

**Acknowledgements**

All research at the Department of Psychiatry in the University of Cambridge is supported by the NIHR Cambridge Biomedical Research Centre (NIHR203312) and the NIHR Applied Research Collaboration East of England. The views expressed are those of the author(s) and not necessarily those of the NIHR or the Department of Health and Social Care.

## Supplementary material

**S1** Compressed association-sensory gradient in schizophrenia at different network density thresholds.

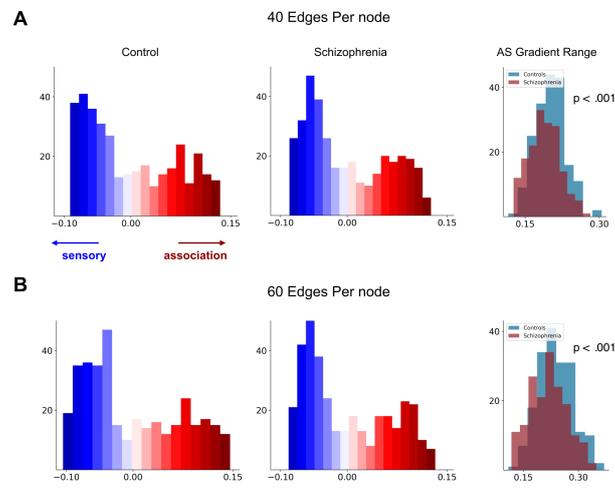

**Fig 1** The compression of AS gradient was consistently observed at sensitivity analyses under 40 (**A**) and 60 (**B**) edges per node.

**S2** Task performance across subtasks.

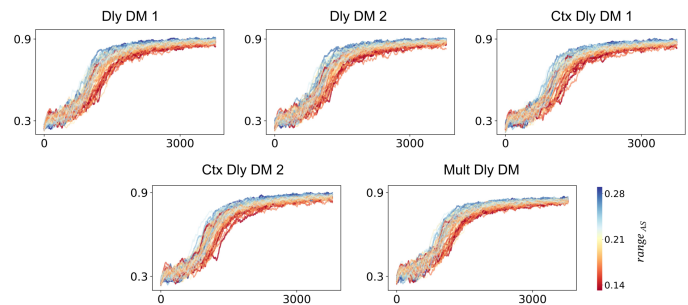

**Fig 2** Across all randomly alternating task variants, networks converged to comparable accuracies. Those regularised with higher association-sensory spectral range tended to rise more quickly and plateau higher.